\documentclass[aps,prd,superscriptaddress,amsfonts,amssymb,amsmath,eqsecnum,nofootinbib,twocolumn,floatfix]{revtex4}
\pdfoutput=1
  \usepackage{color,graphicx}
  \usepackage[utf8]{inputenc}
  \usepackage[T2A]{fontenc}
  \usepackage[english]{babel}
  \definecolor{darkblue}{rgb}{0,0,0.7}
 \definecolor{darkred}{rgb}{0.7,0,0}
 \definecolor{darkgreen}{rgb}{0,0.4,0}
 \usepackage[unicode, colorlinks, citecolor=darkblue, linkcolor=darkred, urlcolor=blue]{hyperref}

 \allowdisplaybreaks
\begin{document}

\author{Sergey P. Vyatchanin}

\affiliation{Faculty of Physics, M.V. Lomonosov Moscow State University, Leninskie Gory, Moscow 119991, Russia}
\affiliation{Quantum Technology Centre, M.V. Lomonosov Moscow State University, Leninskie Gory, Moscow 119991, Russia}

\author{Albert I. Nazmiev}
\affiliation{Faculty of Physics, M.V. Lomonosov Moscow State University, Leninskie Gory, Moscow 119991, Russia}

\author{Andrey B. Matsko}

\affiliation{Jet Propulsion Laboratory, California Institute of Technology, 4800 Oak Grove Drive, Pasadena, California 91109-8099, USA}

\date{\today}
	
\title{Broadband Coherent Multidimensional Variational Measurement}

\begin{abstract}
Standard Quantum Limit (SQL) of a classical mechanical force detection results from quantum back action perturbing evolution of a mechanical system.  In this paper we show that usage of a multidimensional optical transducer may enable a broadband quantum back action evading measurement. We study a corresponding technique for measurement of a resonant signal force acting on a linear mechanical oscillator coupled to an optical system with three optical modes with separation nearly equal to the mechanical frequency. The measurement is performed by optical pumping of the central optical mode and measuring the light escaping the two other modes.  By detecting  optimal quadrature components of the optical modes  and post-processing the measurement results we are able to exclude the back action in a broad frequency band and characterize the force with sensitivity better than SQL. We discuss how proposed scheme relates to multidimensional system containing quantum-mechanics-free subsystems (QMFS) which can evade the SQL using idea of so called ``negative mass'' \cite{TsangPRX2012}.
\end{abstract}

\maketitle

\section{Introduction}

Mechanical motion is frequently observed by usage of optical transducers. These transducers give one a possibility to detect displacement, speed, acceleration, and rotation of mechanical systems. Mechanical motion can change frequency, amplitude and phase of the probe light. The sensitivity of the measurement can be extremely high. For example, a relative mechanical displacement orders of magnitude smaller than a proton size can be detected. This feature is utilized in gravitational wave detectors \cite{aLIGO2013,aLIGO2015,MartynovPRD16,AserneseCQG15, DooleyCQG16,AsoPRD13}, in magnetometers \cite{ForstnerPRL2012, LiOptica2018}, and in torque sensors \cite{WuPRX2014, KimNC2016, AhnNT2020}. 

There are several reasons limiting the fundamental sensitivity of the measurement. One of them is the fundamental thermodynamic fluctuations of the probe mechanical system. The absolute position measurement is restricted due to the Nyquist noise. However, this obstacle can be either decreased or removed if one  measures a variation of the position during time much faster than the system ring down time \cite{Braginsky68, 92BookBrKh}.

Another restriction comes from the quantum noise of the meter. On one hand, the accuracy of the measurements  is limited because of their fundamental quantum fluctuations, represented by the shot noise for the optical probe wave. On the other hand, the sensitivity is impacted by the perturbation of the state of the probe mass due to so called ``back action''.  In the case of optical meter the mechanical perturbation results from fluctuations of the light pressure force. Interplay between these two phenomena leads to a so called standard quantum limit (SQL) \cite{Braginsky68, 92BookBrKh} of the sensitivity. 

The reason for SQL is noncommutativity between the probe noise and the quantum back action noise. In a simple displacement sensor the probe noise is represented by the phase noise of light and the back action noise --- by the amplitude noise of light. The signal is contained in the phase of the probe. The relative phase noise decreases with optical power. The relative back action noise increases with the power. The optimal measurement sensitivity corresponds to SQL.  It is not possible to measure the amplitude noise and subtract it from the measurement result, because of phase and amplitude quantum fluctuations of the same wave do not commute. 

The SQL of a mechanical force, acting on free test mass, can be surpassed in a configuration supporting opto-mechanical velocity measurement  \cite{90BrKhPLA,00a1BrGoKhThPRD}. The limit also can be overcome using opto-mechanical rigidity \cite{99a1BrKhPLA, 01a1KhPLA}. Preparation of the probe light in a nonclassical state \cite{LigoNatPh11, LigoNatPhot13, TsePRL19, AsernesePRL19, YapNatPhot20, YuNature20, CripeNat19} as well as detection of a variation of a strongly perturbed optical quadrature \cite{93a1VyMaJETP, 95a1VyZuPLA, 02a1KiLeMaThVyPRD} curbs the quantum back action and lifts SQL. The SQL can be surpassed with coherent quantum noise cancellation \cite{TsangPRL2010, PolzikAdPh2014, MollerNature2017} as well as compensation using an auxiliary medium with negative nonlinearity \cite{matsko99prl}.
Optimization of the measurement scheme by usage a few optical frequency harmonics as a probe also allows beating the SQL. A dichromatic optical probe may lead to observation of such phenomena as negative radiation pressure \cite{Povinelli05ol,maslov13pra} and optical quadrature-dependent quantum back action evasion \cite{21a1VyNaMaPRA}. 

The first way of back action evading (BAE) for mechanical oscillator, proposed about forty years ago \cite{80a1BrThVo, 81a1BrVoKh}, took advantage of short (stroboscopic) measurements of a  mechanical  coordinate separated by a half period of oscillator. At the same time it was proposed to measure not a coordinate but one of quadrature amplitudes of a mechanical oscillator \cite{Thorne1978, 80a1BrThVo} to perform a BAE. Both propositions are equivalent and can be realized with a pulsing pump \cite{81a1BrVoKh, Clerk08, 18a1VyMaJOSA}.

The measurement proposed here belongs to the class of broadband variational \cite{02a1KiLeMaThVyPRD} measurements for mechanical oscillator. It involves a pump of the main optical mode at frequency $\omega_0$ and  excitation of two additional optical modes at frequencies $\omega_\pm=\omega_0\pm \omega_m$ detuned from pumped mode by eigen frequency of mechanical oscillator $\omega_m$. We propose to detect modes $\omega_\pm$ independently, measuring in each channel an optimal optical quadrature using balance homodyne detection with carriers frequencies $\omega_\pm$. This two channel registration allows us to detect back action  and remove it completely from the measured data.

It was shown recently that a multidimensional quantum system may have subsystems that behave classically \cite{TsangPRX2012}. All the observables of these quantum-mechanics-free subsystem (QMFS) can be used for QND measurements. The noncompatible variables are not coupled in one QMFS and the back action is caused by the observables from another QMFS and, hence, can be removed. We have considered a two mode opto-mechanical read-out and noticed that in some realization of such a system the back action, defined by the sum of quadrature components $(a_++a_+^\dag+a_-+a_-^\dag)/2$ of the modes, impacts the difference of the quadrature components of the modes $(a_++a_+^\dag-a_--a_-^\dag)/2$. These two linear combinations of the quadrature components do not commute and, hence, the measurement is not fundamentally limited by QND. Importantly, to realize a BAE in the scheme one has to post-process a linear combination of the measured quadratures with spectral frequency-dependent {\em complex} coefficients. This type of measurement can be performed if each of the spectral components is detected by a separate homodyne detector. The measurement result is multiplied on the optimal complex parameter, and the results are added together for the force determination.

Our measurement scheme is especially efficient when the signal force is resonant with the mechanical probe mass oscillator. In this regime the external force modifies the power redistribution between the probe spectral components in the optimal way. 

The measurement idea is introduced in Section II using the idealized all-resonant physical model.  The sensitivity of a system with frequency detunings is studied in Section III. It is shown that the non-equidistant modes deteriorate performance of the method. 
In Section IV we relate our scheme  with quantum-mechanics-free subsystem (QMFS) \cite{TsangPRX2012} which can be used for QND measurements.  Section V concludes the paper.


\begin{figure}
 \includegraphics[width=0.4\textwidth]{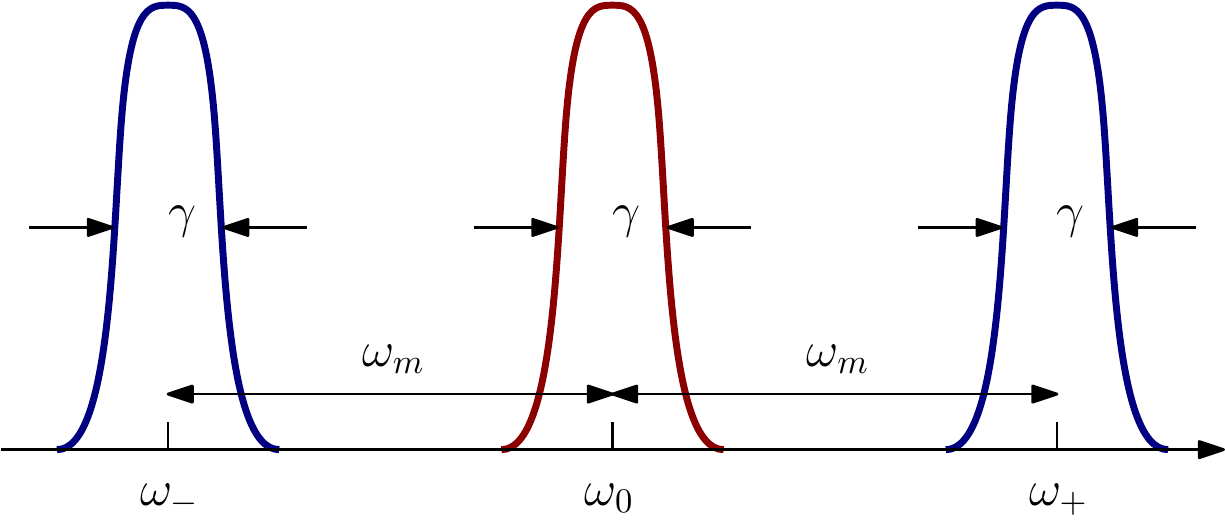}
 \caption{Optomechanical scheme. Three optical modes which frequencies are separated by frequency $\omega_m$ of mechanical oscillator. Optical modes are coupled with the mechanical oscillator. Relaxation rate $\gamma$ is the same for the all tree modes, $\gamma\ll \omega_m$. The middle mode with frequency $\omega_0$ is resonantly pumped. }\label{scheme}
\end{figure}

\section{Physical Model}
\label{Model}

Let we have three optical modes $\omega_-,\ \omega_0,\ \omega_+$ separated from each others by eigen frequency $\omega_m$ of mechanical oscillator as shown on Fig.~\ref{scheme}. The middle mode with frequency $\omega_0$ is resonantly pumped, the modes  $\omega_\pm$ are not pumped. Mechanical oscillator is coupled with optical modes. The photons in the modes are generated due to parametric interaction of the optical pump and the mechanical signal photons. We detect output of modes $\omega_\pm$.

We  assume that the relaxation rates of the optical modes are identical and characterized with the full width at the half maximum (FWHM) equal to $2\gamma$.  The  mechanical relaxation rate $\gamma_m$ is small as compared with the optical one. We also assume that the conditions of the resolved side band interaction and frequency synchronisation are valid:  
\begin{align}
 \label{RSB}
  \gamma_m\ll \gamma \ll \omega_m, \quad \omega_0-\omega_- =\omega_+ -\omega_0=\omega_m
\end{align}

\subsection{Hamiltonian}

The generalized Hamiltonian describing the system can be presented in form
\begin{subequations}
   \label{Halt}
  \begin{align}
  H   &= H_0 + H_\text{int}+H_s+H_T+H_\gamma+H_{T, \, m}+H_{\gamma_ m} ,\nonumber\\
  \label{H0}
  H_0 &=\hslash \omega_+\hat c_+^\dag \hat c_+  + \hslash \omega_0 \hat c_0^\dag \hat c_0 +\\
    &\qquad + \hslash \omega_-\hat c_-^\dag \hat c_-
         +\hslash \omega_m \hat d^\dag \hat d,\nonumber\\
  \label{Hint}
    H_\text{int} & = \frac{\hslash }{i}
      \left(\eta \left[\hat c_0^\dag \hat c_- + \hat c_+^\dag \hat c_0\right] \hat d -\right.\\
     &\qquad -\left. \eta^* \left[\hat c_0 \hat c_-^\dag+ \hat c_+ \hat c_0^\dag \right]\hat  d^\dag \right),\nonumber\\
   \label{Hs}\hat 
   H_s & = - F_s x_0\left(\hat d+ \hat d^\dag\right) 
  \end{align}
  \end{subequations}
$\hat d,\ \hat d^\dag$ are annihilation and creation operators of the mechanical oscillator,  $\hat c_\pm,\ \hat c^\dag_\pm$ are annihilation and creation operators of the corresponding optical modes. The operator of coordinate $x$ of the mechanical oscillator can be presented in form
\begin{align}
\label{x}
 x= x_0\left(\hat d + \hat d^\dag\right),\quad x_0=\sqrt\frac{\hslash}{2m \omega_m}.
\end{align}
$H_\text{int}$ is the Hamiltonian of the interaction between optical and mechanical modes (fast oscillating terms are omitted), $\eta$ is coupling constant. $H_s$ is Hamiltonian of signal force $F_s$.
$H_T$ is the Hamiltonian describing the environment (thermal bath) and  $H_\gamma$ is the Hamiltonian of the coupling between the environment and the optical modes resulting in decay rate $\gamma$. We neglect by the internal loss in the optical system. The pump is also included into $H_\gamma$. Similarly, $H_{T, \, m}$ is the Hamiltonian of the environment and $H_{\gamma_m}$ is the Hamiltonian describing coupling between the environment and the mechanical oscillator resulting in decay rate $\gamma_m$. See Appendix~\ref{IntrDeriv} for details.

We denote the normalized input and output optical amplitudes as $\hat a_{\pm, \,0}$ and  $\hat b_{\pm, \,0}$ correspondingly. Using the Hamiltonian \eqref{Halt} we derive the equations of motion for the intracavity fields (see Appendix~\ref{IntrDeriv} for the full derivation).
\begin{subequations}
\begin{align} 
\dot {\hat c}_0+\gamma \hat c_0&=\eta^*\hat c_+ \hat d^\dag - \eta \hat c_- \hat d +\sqrt{2 \gamma}\,\hat a_0,\\
\dot { \hat c}_-+\gamma \hat c_-&=\eta^*\hat c_0 \hat d^\dag +\sqrt{2 \gamma}\,\hat a_m,\\
\dot {\hat c}_++\gamma \hat c_+&=-\eta \hat c_0 \hat d  +\sqrt{2 \gamma}\,\hat a_p,\\
\dot {\hat d}+\gamma_m \hat d&=\eta^*(\hat c_0 \hat c_-^\dag + \hat c_0^\dag \hat c_+)+\sqrt{2 \gamma_m}\,\hat q +f_s.
\end{align}  \label{moveq}
\end{subequations} 
Here $\hat q$ is fluctuation force acting on mechanical oscillator, and $f_s$ is normalized signal force (see definition \eqref{fs} below).

The input-output relations connecting the external and intracavity fields are
\begin{align}
 \label{outputT}
  \hat b_\pm= -\hat a_\pm + \sqrt{2\gamma} \hat c_\pm.
\end{align} 

It is convenient to separate the expectation values of the wave amplitudes at frequency $\omega_0$ (described by block letters) as well as its fluctuation part (described by small letters) and assume that the fluctuations are small:
 \begin{align}
 \label{expA}
 \hat c_0 & \Rightarrow C_0 e^{-i\omega_0 t}+  c_0 e^{-i\omega_0 t},
 \end{align}
$C_0$  stands for the expectation value of the field amplitude in the mode with eigenfrequency $\omega_0$ and  $c_{  0}$ represent the fluctuations of the field in the mode, $|C_0|^2 \gg \langle \hat c_0^\dag \hat c_0 \rangle$, where $\langle \dots \rangle$ stands for ensemble averaging. Similar expressions can be written for the optical modes with eigenfrequencies  $\omega_{\pm}$ and the mechanical mode  with eigenfrequency $\omega_m$. The normalization of the amplitudes is selected so that $\hslash \omega_0 |A_0|^2$ describes the optical power \cite{02a1KiLeMaThVyPRD}. 

Using the equations of motion \eqref{moveq} and  assuming that $A_+=A_-=0$ (the regular signal contribution is considered in the fluctational parts) we get the (zeroth order of approximation) equations for the expectation values
\begin{subequations}
\begin{align} 
 \gamma C_0&= \eta^* C_+ D^*-\eta C_- D +\sqrt{2 \gamma}A_0,\\
\gamma C_-&=  \eta^* C_0  D^* ,\\
\gamma C_+&=-  \eta C_0 D ,\\
\gamma_m D&= \eta^*(C C_-^*  +  C^* C_+) .
\end{align}  
  \end{subequations} 
This set of equations has an obvious stationary solution
\begin{equation}
C_0=\sqrt{\frac{2}{\gamma}}A_0, \quad C_-=C_+=0. \label{sol}
\end{equation}

Substituting Eq.~(\ref{sol}) into the equations of motion \eqref{moveq} we derive the stability conditions for this solution.
\begin{subequations}
\begin{align} 
\dot c_0+\gamma c_0&= 0 \label{eq1},\\
\dot c_-+\gamma c_-&=  \eta^*C_0 d^* ,\\
\dot c_++\gamma c_+&=-\eta C_0 d ,\\
\dot d+\gamma_m d&=  \eta^*( c_-^* C_0+ c_+C_0^*).
\end{align}  
  \end{subequations} 
The first equation \eqref{eq1} for the middle mode  separates from the other three. 

Substituting $c_+=c_+ e^{\lambda t}$, $c_-=c_- e^{\lambda t}$, $d=d e^{\lambda t}$ we get
\begin{equation}
\begin{matrix}
(\lambda+\gamma) c_-^* & + &0 \cdot c_+ & - &  \eta  C_0^*d & =0,\\
0\cdot c_-^* & +&(\lambda+\gamma)  c_+ & + & \eta  C_0 d & =0,\\
   -  \eta^* C_0c_-^* & -&\eta^* C_0^*c_+ & -&(\lambda+\gamma_m) d & =0.
\end{matrix}
\end{equation}

Solving equation $\Delta=0$, where $\Delta$ is the determinant of this set of linear equations, we get $\lambda_{1,2}=-\gamma<0$, $\lambda_3=-\gamma_m<0$, hence the solution of the linearized equations \eqref{sol} is stable. 

\begin{figure}
 \includegraphics[width=0.45\textwidth]{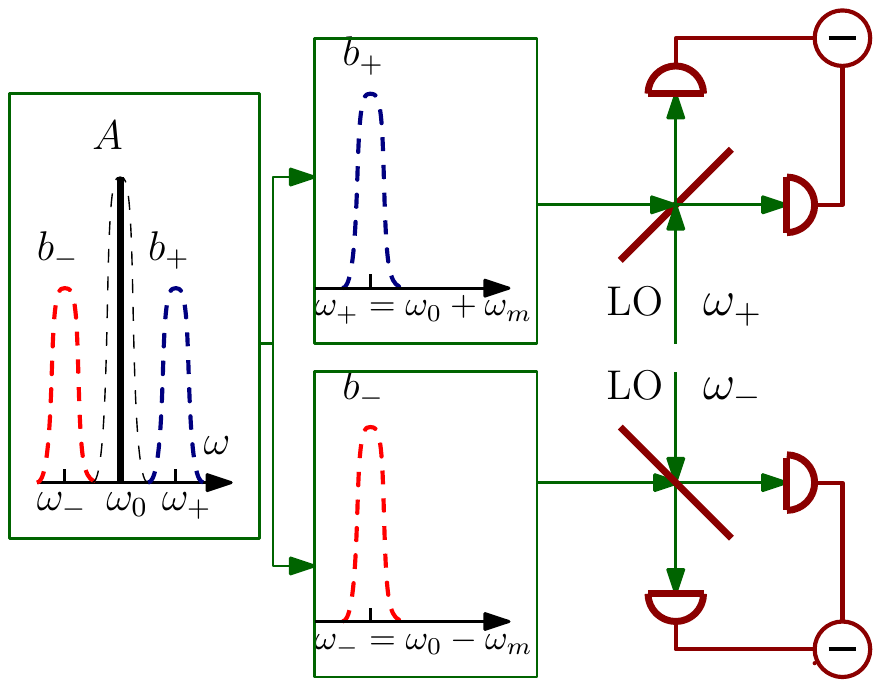}
 \caption{A scheme of the measurement. Quadrature components of the output modes $\omega_\pm$ are measured separately by balanced homodyne detectors with corresponding optimal local oscillators having frequencies $\omega_\pm$. The signal is inferred by processing of the linear combination of the measured results. Essentially, the linear combination in frequency domain should have complex frequency dependent coefficients. }\label{scheme2}
\end{figure}

Substituting the solution \eqref{sol} into the equations of motion \eqref{moveq} we finally obtain 
\begin{subequations}
\begin{align}
\label{c0}
\dot {\hat c}_0+\gamma \hat c_0&= \sqrt{2 \gamma}\hat a_0,\\
 \label{c+T}
  \dot {\hat   c}_+  + \gamma \hat c_+ + \eta C_0 \hat d &= \sqrt {2 \gamma} \hat a_+,\\
  \label{c-T}
  \dot {\hat   c}_-  + \gamma \hat c_- - \eta^* C_0^* \hat d^\dag &= \sqrt {2 \gamma} \hat a_-,\\
  \dot {\hat d} +\gamma_m \hat d 
    - \eta^* \left[C_0 \hat c_-^\dag+ \hat c_+ C_0^* \right] &=\sqrt{2\gamma_m}\hat q + f_s
\end{align}
\end{subequations}
Outputs $b_\pm$ around frequencies $\omega_\pm$ have to be detected separately, as shown in Fig.~\ref{scheme2}. We also see that fluctuation waves around $\omega_0$ do not influence on field components in the vicinity of frequencies $\omega_\pm$ and the first equation \eqref{c0} separates from the other three, so it is omitted in the further consideration. 

We assume in what follows that the expectation amplitudes are real, same as the coupling constant $\eta$
\begin{align}
 \label{real}
 C_0= C_0^*,\quad A_0=A_0^*,\quad \eta=\eta^*
\end{align}

The operators $\hat a_\pm$ are characterized with the following commutators and correlators
\begin{align}
  \label{comm}
  \left[\hat a_\pm(t), \hat a_\pm^\dag(t')\right] &=   \delta(t-t'),\\
  \label{corr}
  \left\langle\hat a_\pm(t) \hat a_\pm^\dag(t')\right\rangle &= \delta(t-t'),
\end{align}
where $\langle \dots \rangle$ stands for ensemble averaging. This is true since the incident fields are considered to be in the coherent state. 

The Fourier transform of these operators is defined as follows
\begin{align}
 \hat a_\pm (t) &= \int_{-\infty}^\infty a_\pm(\Omega) \, e^{-i\Omega t}\, \frac{d\Omega}{2\pi}.
\end{align}
Similar expressions can be written for the other operators. Using \eqref{comm} and \eqref{corr} we derive commutators and correlators for the Fourier amplitudes of the input fluctuation operators
 \begin{align}
  \label{comm1}
  \left[ a_\pm(\Omega),  a_\pm^\dag(\Omega')\right] &= 2\pi\,\delta(\Omega -\Omega'),\\
  \label{corr1}
  \left\langle a_\pm(\Omega)  a_\pm^\dag(\Omega')\right\rangle &= 2\pi\, \delta(\Omega -\Omega')
\end{align}
\subsection{Solution} \label{Solution}

The Fourier amplitudes  for the the intracavity field as well as mechanical amplitude, $c_\pm$ and $d$, can be found utilizing (\ref{c+T}, \ref{c-T}):
\begin{subequations}
 \label{set1}
\begin{align} 
\label{c+}   
 (\gamma - i\Omega)c_+(\Omega) & +  \eta C_0 d(\Omega) = \sqrt {2 \gamma}  { a}_{+}(\Omega),\\
\label{c-} 
 (\gamma - i\Omega) c_-(\Omega) & - \eta C_0 d^\dag(-\Omega)  = \sqrt {2\gamma} a_-(\Omega),\\
\label{dd}
 (\gamma_m - i\Omega) d(\Omega) -  &    \eta C_0  \Big(c_+(\Omega)+ c_-^\dag(-\Omega)\Big) =\\
 \nonumber
    & =  \sqrt{2\gamma_m}\, q(\Omega)+      f_s(\Omega)\\ 
\label{output}
  b_{ \pm}(\Omega) &= - { a}_{\pm}(\Omega) + \sqrt {2 \gamma}\, { c}_\pm(\Omega),
\end{align} 
\end{subequations}
We assume that the signal force is a resonant square pulse acting during time interval $\tau$:
\begin{align}
\label{Fs}
F_S(t)&= F_{s0}\sin(\omega_m t + \psi_f) = \\
    = &i(F_{s}(t) e^{-i\omega_m t} - F_{s}^*(t) e^{i\omega_m t}),\quad 
  -\frac{\tau}{2} < t < \frac{\tau}{2},\nonumber\\
  \label{fs}
  f_s(\Omega)& = \frac{F_s(\Omega)}{\sqrt{2\hslash \omega_m m}},\quad 
  f_{s0} = \frac{F_{s0}(\Omega)}{\sqrt{2\hslash \omega_m m}}= 2f_s.
\end{align}
where  $F_s(\Omega)\ne F_s^*(-\Omega)$ is the Fourier amplitude of $F_s(t)$.  

The Fourier amplitudes of the thermal noise operators $\hat q$ obey to the relations
\begin{subequations}
 \label{qDef}
\begin{align}
 \label{commq}
  \left[ q(\Omega),\,  q^\dag(\Omega')\right] &= 2\pi\,  \delta(\Omega-\Omega'),\\
  \label{corrq}
  \left\langle q(\Omega)\,  q^\dag(\Omega')\right\rangle 
    &= 2\pi\,  \big(2n_T+1\big)\, \delta(\Omega-\Omega'),\\
    n_T&= \dfrac{1}{1 - e^{-\hslash \omega_m/\kappa_BT} },
\end{align}
\end{subequations}
where $\kappa_B$ is Boltzmann constant, $T$ is the ambient  temperature.

Introducing quadrature amplitudes of amplitude and phase
\begin{subequations}
\label{quadDef}
 \begin{align}
  a_{\pm a} &= \frac{a_\pm (\Omega) +a_\pm ^\dag(-\Omega)}{\sqrt 2}\,,\\
	 a_{\pm \phi} &= \frac{a_\pm (\Omega) -a_\pm ^\dag(-\Omega)}{i\sqrt 2}\,.
 \end{align}
\end{subequations}
(the quadrature amplitudes for the other operators are introduced in the same way)
and using \eqref{set1} we obtain
\begin{subequations}
  \label{quadIn}
 \begin{align}
  \label{ca+}   
 (\gamma - i\Omega)c_{+a} +  \eta C_0  d_a &= \sqrt {2 \gamma}  { a}_{ +a},\\
 \label{cphi+}   
 (\gamma - i\Omega)c_{+\phi} +  \eta C_0  d_\phi &= \sqrt {2 \gamma}  { a}_{ +\phi},\\
\label{ca-} 
 (\gamma - i\Omega) c_{-a} - \eta C_0  d_a &= \sqrt {2 \gamma}  { a}_{-a},\\
\label{cphi-} 
 (\gamma - i\Omega) c_{-\phi} + \eta  C_0  d_\phi &= \sqrt {2 \gamma}  { a}_{-\phi},\\
\label{da}
 (\gamma_m - i\Omega)  d_a - \eta C_0 &\Big(c_{+a}+c_{-a}\Big) =\\
        &= \sqrt {2 \gamma_m} q_a +  f_{s\,a},\nonumber\\
\label{dphi}
 (\gamma_m - i\Omega)  d_\phi  - \eta C_0 & \Big(c_{+\phi} - c_{-\phi}\Big) =\\
        &= \sqrt {2 \gamma_m} q_{\phi} +  f_{s\,\phi}.\nonumber
 \end{align}
 \end{subequations}
Please note that sum $c_{+a} +c_{-a}$ does not contain  information on the mechanical motion (term proportional to $\sim d_a$ is absent), but produces the back action term in \eqref{da}. Introducing sum and difference of the quadratures
\begin{align}
 \label{gDef}
 g_{a\pm} &= \frac{c_{+a}\pm c_{-a}}{\sqrt 2},\quad 
  g_{\phi\pm} = \frac{c_{+\phi}\pm c_{-\phi}}{\sqrt 2},\\
 \label{alphaDef}
 \alpha_{a\pm}&= \frac{a_{+a}\pm a_{-a}}{\sqrt 2}\,,\quad
    \alpha_{\phi\pm}= \frac{a_{+\phi}\pm a_{-\phi}}{\sqrt 2},\\
 \label{betaDef}
 \beta_{a\pm}&= \frac{b_{+a}\pm b_{-a}}{\sqrt 2}\,,\quad
    \beta_{\phi\pm}= \frac{b_{+\phi}\pm b_{-\phi}}{\sqrt 2}
\end{align}
and rewriting \eqref{quadIn} in the new notations we obtain
\begin{subequations}
  \label{quadIn2}
 \begin{align}
  \label{ga+}   
 (\gamma - i\Omega)g_{a+} = \sqrt {2 \gamma}  \alpha_{ a+},\\
 \label{ga-} 
 (\gamma - i\Omega) g_{a-} + \sqrt 2 \eta C_0  d_a &= \sqrt {2 \gamma}  \alpha_{a-},\\
 \label{da2}
 (\gamma_m - i\Omega)  d_a - \sqrt 2 \eta C_0 g_{a+} 
        &= \sqrt {2 \gamma_m} q_a +  f_{s\,a},\\
 \label{gphi+}   
 (\gamma - i\Omega)g_{\phi+} +  \sqrt 2\eta C_0  d_\phi &= \sqrt {2 \gamma}  \alpha_{ \phi+},\\
\label{gphi-} 
 (\gamma - i\Omega) g_{\phi-}  &= \sqrt {2 \gamma} \alpha_{\phi-},\\
\label{dphi2}
(\gamma_m - i\Omega)  d_\phi  - \sqrt 2 \eta C_0g_{\phi-} 
        &= \sqrt {2 \gamma_m} q_{\phi} +  f_{s\,\phi}.
 \end{align}
 \end{subequations}
The sets (\ref{ga+}, \ref{ga-}, \ref{da2}) and (\ref{gphi+}, \ref{gphi-}, \ref{dphi2}) are independent and can be separated.

It is convenient to present the solution of set (\ref{ga+}, \ref{ga-}, \ref{da2}) for the amplitude quadratures in form
\begin{subequations}
 \label{betaMSIa}
 \begin{align}
  \beta_{+a}   &= \xi\, \alpha_{+a},\quad \xi=\frac{\gamma+i\Omega}{\gamma-i\Omega},\\
  \label{beta-a}
  \beta_{-a}  &=\xi\left(\alpha_{-a} - \frac{\mathcal K\, \alpha_{+a}}{\gamma_m-i\Omega}\right)-\\
    &\qquad         - \frac{\sqrt{\xi \mathcal K }}{\gamma_m-i\Omega}
            \left(\sqrt {2 \gamma_m} q_a + f_{s\,a}\right),\\
        &\quad  \mathcal K\equiv \frac{4  \gamma\,\eta^2 C_0^2}{\gamma^2+\Omega^2}
 \end{align}
 \end{subequations}
As expected, in Eq.~\eqref{beta-a} the back action term is proportional to the normalized probe power $\mathcal K$. However, this term can be excluded by the post processing.
One  can measure {\em both}  $\beta_{+a}$ and $\beta_{-a}$ simultaneously and subtract $\beta_{+a}$ from  $\beta_{-a}$ to remove the back action completely. It means that we can measure combination
 \begin{align}
  \beta_{-a}^\text{comb} &= \beta_{-a} + \xi \, \frac{\mathcal K\, \alpha_{+a}}{\gamma_m-i\Omega} =\\
  \label{beta-a2}
         &=\xi\alpha_{-a} - \frac{\sqrt{\xi \mathcal K }}{\gamma_m-i\Omega}
            \left(\sqrt {2 \gamma_m} q_a + f_{s\,a}\right),
 \end{align}
 which is back action free. This is the main finding of the study. Essentially, the coefficient needed for suppression of the back action is complex. It depends on the spectral frequency $\Omega$. While a similar result was obtained earlier \cite{21a1VyNaMaPRA}, the measurement scheme considered here involves single probe beam and is stable. It does not use the dichromatic pump introducing resonant mechanical motion that has to be controlled.    
 
We find the force detection condition using  single-sided power spectral density $S_{f}(\Omega)$ for signal force \eqref{Fs}. Assuming that the detection limit corresponds to the signal-to-noise ratio exceeding unity we obtain
 \begin{align}
  \label{SfDef}
  f_{s0} \ge \sqrt{S_{f}(\Omega)\cdot \frac{\Delta\Omega}{2\pi} },
 \end{align}
 where $\Delta \Omega \simeq 2\pi/\tau$. 
 Using (\ref{corr1}, \ref{corrq}) we derive for the case when we measure  $ \beta_{-a}$ \eqref{beta-a}
 \begin{align}
  \label{Sf}
  S_{f}(\Omega) &= 2\gamma_m\big(2n_T+1\big) 
    + \frac{\gamma_m^2+\Omega^2}{\mathcal K} +\mathcal K\ge\\
    \label{SQL}
    &\ge 2\gamma_m\big(2n_T+1\big) +S_{SQL,f},\\ 
  S_{SQL,f} &= 2\sqrt{\gamma_m^2+\Omega^2}\label{SSQL}
 \end{align}
The sensitivity is restricted by SQL.
If me measure $\beta_{-a}^\text{comb}$ \eqref{beta-a2} the spectral density is not limited by SQL
 \begin{align}
  \label{Sfb}
  S_{f}(\Omega) &= 2\gamma_m\big(2n_T+1\big) + \frac{\gamma_m^2+\Omega^2}{\mathcal K}
 \end{align}
Here the first term describes thermal noise and the second one stands for the quantum measurement noise (shot noise) decreasing with the power increase. The back action term is excluded completely. 

The thermal noise masks signals in any opto-mechanical detection scheme. It cannot be separated from the signal if it comes in the same channel as the force, at the same time, and with spectral components overlapping with the signal. The error associated with the thermal noise can exceed the measurement error related to the measurement system. A proper measurement procedure allows to reduce the impact of the thermal noise not identical to the signal force and coming from the apparatus itself and also exclude the quantum uncertainty associated with the initial state of the mechanical system. The main requirement for such a measurement is fast interrogation time $\tau$, which should be much shorter than the ring down time of the mechanical system, i.e. $\gamma_m\tau\ll 1$ \cite{Braginsky68,92BookBrKh}. This is possible if the measurement bandwidth exceeds the bandwidth of the mechanical mode. Sensitivity of narrowband resonant measurements is usually limited by the thermal noise.
 
One can measure sum and differences of the phase quadratures instead of the amplitude quadratures. Solving set (\ref{gphi+}, \ref{gphi-}, \ref{dphi2}) we arrive at
 \begin{subequations}
  \label{betaMSIphi}
 \begin{align}
  \beta_{-\phi} &= \xi\, \alpha_{-\phi},\\
  \label{beta-phi}
  \beta_{+\phi}  &=\xi\left(\alpha_{+\phi} - \frac{\mathcal K\, \alpha_{-\phi}}{\gamma_m-i\Omega}\right)-\\
    &\qquad         - \frac{\sqrt{\xi \mathcal K }}{\gamma_m-i\Omega}
            \left(\sqrt {2 \gamma_m} q_\phi +  f_{s\,\phi}\right).
 \end{align}
\end{subequations}
We can measure quadratures $\beta_{\pm\phi}$ simultaneously and subtract back action proportional to $\beta_{-\phi}$ from $\beta_{+\phi}$.

A generalization is possible for a pair of quadrature components with arbitrary parameter $\varphi$
\begin{subequations}
 \label{quadphi}
\begin{align}
 b_{+\varphi} &= b_{+a}\cos\varphi +b_{+\phi}\sin\varphi,\\
   b_{-\varphi}&= b_{-a}\cos\varphi -b_{-\phi}\sin\varphi
\end{align}
\end{subequations}
The sum $b_{+\varphi}+ b_{-\varphi}$ is not disturbed by the mechanical motion but contains the term proportional to the back action force, whereas the difference $b_{+\varphi}- b_{-\varphi}$ contains the term proportional to mechanical motion (with back action and signal). The back action term can be measured and subtracted from the force measurement result.

\section{Influence of the detuning}
\label{Detuning}

\begin{figure}
\includegraphics[width=0.4\textwidth]{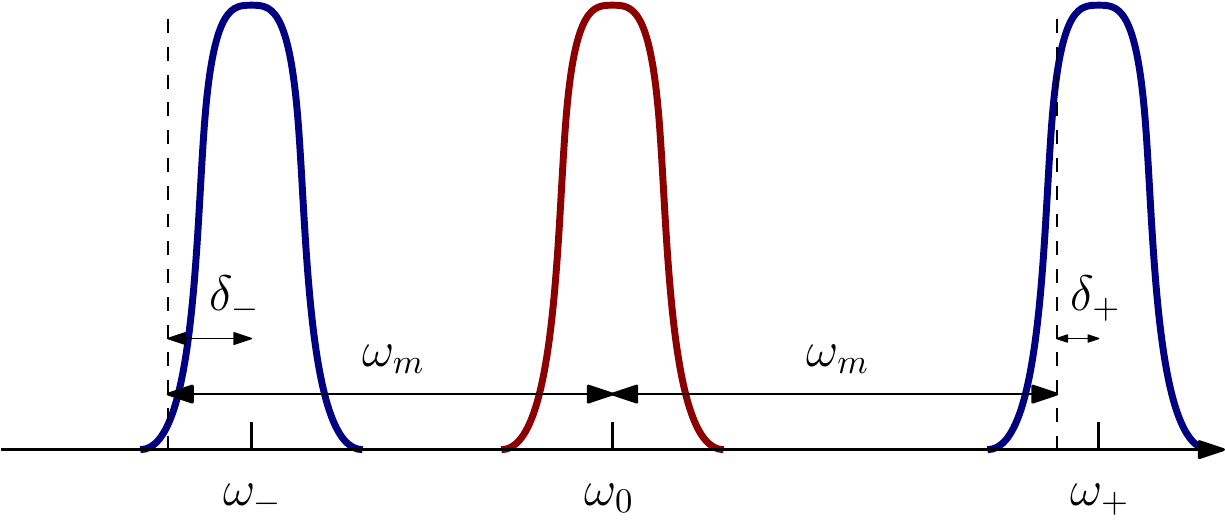}
 \caption{ Frequency chart. General case.}\label{det3}
\end{figure}

 Analysis  in Sec. \ref{Solution} was made under the assumption that the difference between the frequencies of the consecutive optical modes is precisely equal to the mechanical frequency. In this Section we analyze the system characterized with imperfect frequency synchronization conditions. Let us consider frequencies of the optical modes to be shifted by the arbitrary   values $\delta_-$ for the left sideband and $\delta_+$ for the right as shown on Fig.~\ref{det3}.
\begin{equation}
\omega -\omega_- = \omega_m-\delta_-,\qquad \omega_+-\omega = \omega_m+\delta_+.
\end{equation}
The nonzero frequency detuning $\delta_0$ of the pump light from the resonant frequency $\omega_0$ and frequency detuning $\delta_f$ of the signal force from the mechanical frequency $\omega_m$ can be neglected here. The pump frequency can be locked to the resonator mode. We also expect that the dimensions of the optical resonator can be adjusted so that $\delta_f=0$.  The simplified equations of motions take the following form
\begin{subequations}
\begin{align} 
\dot {\hat c}_0+\gamma \hat c_0 &=\eta^*\hat c_+ \hat d^\dag - \eta \hat c_- \hat d +\sqrt{2 \gamma}\hat a_0, \label{delta1} \\
\dot { \hat c}_-+(\gamma-i\delta_-) \hat c_-& =\eta^*\hat c_0 \hat d^\dag +\sqrt{2 \gamma}\hat a_-,\label{delta2}\\
\dot {\hat c}_++(\gamma-i\delta_+) \hat c_+ &=-\eta \hat c_0 \hat d  +\sqrt{2 \gamma}\hat a_+,\label{delta3}\\
\dot {\hat d}+\gamma_m \hat d =\eta^*&(\hat c_0 \hat c_-^\dag + \hat c_0^\dag \hat c_+)+\sqrt{2 \gamma_m}\hat q +f_s. 
\end{align}  \label{moveq4b}
  \end{subequations} 
The detuning values are considered to be small in comparison with the spectral width of the optical modes $\delta_\pm \ll \gamma$. The difference between \eqref{moveq} and \eqref{moveq4b} is in the presence of $\ i\delta_\pm \hat c_\pm$ terms in the second and third equations (\ref{delta2}, \ref{delta3}).

The expectation values for the amplitudes of the optical and mechanical modes is 
\begin{equation}
C_0=\sqrt{\frac{2}{\gamma}}A_0, \qquad C_-=C_+=D=0.
\end{equation}
The solution is stable as the roots of the characteristic equation are real and negative $\Re\lambda_{1,2}=-\gamma<0$, $\lambda_3=- \gamma_m $. 

For the sake of simplicity and consistency we assume that coupling constant $\eta$  and  the mean amplitude $C_0$ of the intracavity field are real
\begin{equation}
    C_0=C_0^*, \quad \eta=\eta^*.
\end{equation}

Substituting the expectation values of the amplitudes into \eqref{moveq4b} we derive the equations of motion for the Fourier amplitudes of the fluctuation parts of the optical sidebands and the the mechanical oscillation
\begin{subequations}
\begin{align} 
(\gamma-i\delta_--i \Omega) c_-&= \eta C_0  d^\dag+\sqrt{2 \gamma}a_-,\\
(\gamma-i\delta_+-i \Omega) c_+&=- \eta C_0 d +\sqrt{2 \gamma}a_+,\\
(\gamma_m - i \Omega)  d = \eta  C_0 & (  c_-^\dag +   c_+) +\sqrt{2 \gamma_m}q+f_s .
\end{align}  \label{moveq33}
  \end{subequations}  

Introducing the quadratures in the same way as we did it in \eqref{quadDef}, 
we obtain equations for the quadratures. We denote them with apostrophes to distinguish from  \eqref{quadIn}
\begin{subequations}
\begin{align}
\begin{split}
    c_{+a}'=&\frac{(-\eta C_0 d_a+\sqrt{2\gamma} a_{+a})(\gamma-i\Omega)}{(\gamma-i\Omega)^2+\delta_+^2}-\\
   &-\frac{\delta_+(-\eta C_0d_\phi+\sqrt{2\gamma}a_{+\phi})}{(\gamma-i\Omega)^2+\delta_+^2},
\end{split}\\
\begin{split}
    c_{-a}'=&\frac{(\eta C_0 d_a+\sqrt{2\gamma} a_{-a})(\gamma-i\Omega)}{(\gamma-i\Omega)^2+\delta_-^2}-\\
    &-\frac{\delta_-(-\eta C_0d_\phi+\sqrt{2\gamma}a_{-\phi})}{(\gamma-i\Omega)^2+\delta_-^2},
\end{split}\\
    \begin{split}
        c_{+\phi}'=&\frac{(-\eta C_0 d_\phi+\sqrt{2\gamma} a_{+\phi})(\gamma-i\Omega)}{(\gamma-i\Omega)^2+\delta_+^2}+\\
    &+\frac{\delta_+(-\eta C_0d_a+\sqrt{2\gamma}a_{+a})}{(\gamma-i\Omega)^2+\delta_+^2},
    \end{split}\\
   \begin{split}
      c_{-\phi}'= &\frac{(-\eta C_0 d_\phi+\sqrt{2\gamma} a_{-\phi})(\gamma-i\Omega)}{(\gamma-i\Omega)^2+\delta_-^2}+\\
   &+\frac{\delta_-(\eta C_0d_a+\sqrt{2\gamma}a_{-a})}{(\gamma-i\Omega)^2+\delta_-^2}  ,
   \end{split} \\
   \begin{split}
    (\gamma_m - i\Omega) &  d_a - \eta C_0 \Big(c_{+a}+c_{-a}\Big) =\\
        &= \sqrt {2 \gamma_m} q_a +  f_{s\,a},
   \end{split}\\
   \begin{split}
    (\gamma_m - i\Omega) & d_\phi  - \eta C_0  \Big(c_{+\phi} - c_{-\phi}\Big) =\\
        &= \sqrt {2 \gamma_m} q_{\phi} +  f_{s\,\phi}.
   \end{split}
\end{align} \label{quadNew1}
\end{subequations}
The ``new'' quadratures \eqref{quadNew1} can be expressed as the linear combinations of  ``old'' quadratures.  Saving only terms linear over detuning values $\delta_\pm$ in \eqref{quadNew1} we obtain
\begin{subequations}
\begin{align}
    c_{+a}' &=c_{+a}-\frac{\delta_+}{\gamma-i \Omega}c_{+\phi},\\
    c_{-a}' &=c_{-a}-\frac{\delta_-}{\gamma-i \Omega}c_{-\phi},\\
    c_{+\phi}' &=c_{+\phi}+\frac{\delta_+}{\gamma-i \Omega}c_{+a},\\
    c_{-\phi}' &=c_{-\phi}+\frac{\delta_-}{\gamma-i \Omega}c_{-a},\\
   \begin{split}
    (\gamma_m - i\Omega)  d_a &- \eta C_0 \Big(c_{+a}+c_{-a}\Big) =\\
        &= \sqrt {2 \gamma_m} q_a +  f_{s\,a},
   \end{split}\\
   \begin{split}
    (\gamma_m - i\Omega)  d_\phi & - \eta C_0  \Big(c_{+\phi} - c_{-\phi}\Big) =\\
        &= \sqrt {2 \gamma_m} q_{\phi} +  f_{s\,\phi}.
   \end{split}
\end{align} \label{quadNew}
\end{subequations}
We introduce sum and difference of the quadratures as in \eqref{gDef} and save the terms proportional to the first order of $\delta_\pm$. It is also convenient to introduce symmetric and anti-symmetric combinations of the detunings
\begin{align}
&\Delta=\frac{\delta_++\delta_-}{2}, & &\delta=\frac{\delta_+-\delta_-}{2}.
\end{align}
We denote sum and difference of the ``new"''quadratures by the apostrophes, we express them as the liner combinations of the ``old'' \eqref{gDef} expressions
\begin{subequations}
\label{g'}
\begin{align}
& g_{a+}'=g_{a+}-\frac{\delta}{\gamma-i\Omega}g_{\phi-}-\frac{\Delta}{\gamma-i\Omega}g_{\phi+},\\
&g_{a-}'=g_{a-}-\frac{\Delta}{\gamma-i\Omega}g_{\phi-}-\frac{\delta}{\gamma-i\Omega}g_{\phi+},\\
&(\gamma_m-i \Omega)d_a= \sqrt{2}\eta C_0 g_{a+}+\sqrt{2\gamma_m}q_a+f_a ,\\ 
&g_{\phi-}'=g_{\phi-}+\frac{\delta}{\gamma-i\Omega}g_{a+}+\frac{\Delta}{\gamma-i\Omega}g_{a-},\\
& g_{\phi+}'=g_{\phi+}+\frac{\Delta}{\gamma-i\Omega}g_{a+}+\frac{\delta}{\gamma-i\Omega}g_{a-},\\
&(\gamma_m-i \Omega)d_\phi= \sqrt{2}\eta C_0 g_{\phi-}+\sqrt{2\gamma_m}q_\phi+f_\phi.
\end{align}  
  \end{subequations} 

The sum and difference of the amplitude quadratures of the output fields can be obtained using the input-output relations \eqref{outputT}:
%
%
%
\begin{subequations}
\begin{align}
&\beta_{a+}'=\beta_{a+}-\frac{\delta\sqrt{2\gamma}}{\gamma-i\Omega}g_{\phi-}-\frac{\Delta\sqrt{2\gamma}}{\gamma-i\Omega}g_{\phi+},\\
&\beta_{a-}'=\beta_{a-}-\frac{\Delta\sqrt{2\gamma}}{\gamma-i\Omega}g_{\phi-}-\frac{\delta\sqrt{2\gamma}}{\gamma-i\Omega}g_{\phi+},
\end{align}
  \end{subequations} 

The measurement of the combination \eqref{beta-a2}  suppresses the major part of back-action, same as in the perfectly tuned case. However, the parts, proportional to $\delta$ and $\Delta$ cannot be removed.
\begin{subequations}
\begin{align}
    {\beta_a^{comb}}'&=\frac{\mathcal{K}}{\gamma_m-i\Omega}\beta_{+a}'+\beta_{-a}'\approx\\
  \approx &\xi\alpha_{-a}
   -\frac{\sqrt{\xi \mathcal K }}{\gamma_m-i\Omega}
   \left(\sqrt {2 \gamma_m} q_a + f_{s\,a}\right)- \nonumber\\
   \begin{split}
 -& \frac{\mathcal{K}\, }{(\gamma_m-i\Omega)(\gamma-i\Omega)}
    \left[\delta -\frac{\Delta \xi\mathcal K}{(\gamma_m-i\Omega)}\right]
    \alpha_{\phi-}\\
    &\quad-\frac{2\gamma\Delta\mathcal{K}}{(\gamma-i\Omega)^2(\gamma_m-i\Omega)}\alpha_{ \phi+}-
    \end{split}\\
    &   \quad + \frac{\sqrt{\xi \mathcal K }\Delta \mathcal K}{(\gamma-i\Omega)(\gamma_m-i\Omega)^2}
  \left(\sqrt {2 \gamma_m} q_\phi + f_{s\,\phi}\right). \nonumber
\end{align} \label{betaa}
\end{subequations}
In this case it is optimal to measure  but combination of the quadrature components of the force

\begin{equation}
    f_{s}=\cfrac{f_{s,\, a}+D\mathcal{K} f_{s,\,\phi}}{\sqrt{1+\left|D\mathcal{K} \right|^2}},
\end{equation}
where 
\begin{equation}
     D= \frac{\Delta}{(\gamma-i\Omega)(\gamma_m-i\Omega)}.
\end{equation}
The noise power spectral density calculated from \eqref{betaa} takes form
\begin{align}
 S(\Omega) &=2\gamma_m(n_T+1) +\cfrac{\gamma_m^2+\Omega^2}{\mathcal{K}(1+|D|^2\mathcal K^2)} \label{Sdetb}\\ \nonumber
    +&  \frac{\mathcal{K}
    \left [ \left|\delta - \xi\mathcal K D(\gamma-i\Omega)\right|^2
    + 4\gamma^2|D|^2 (\gamma_m^2+\Omega^2) \right] }{(\gamma^2+\Omega^2)(1+|D|^2\mathcal K^2)} .\nonumber
\end{align}
It is reasonable to analyze three values of pump level. For the small enough pump we get
\begin{subequations}
\begin{align}
  \mathcal K  |D| &\ll \left|\frac{\delta}{\gamma-i\Omega}\right|,\\ 
 &\quad \text{or}\  \mathcal{K}\ll \mathcal{K}_{crit1}=\frac{\sqrt{\gamma_m^2+\Omega^2}|\delta|}{|\Delta|}. \label{cond_small}
 \end{align}
\end{subequations}
In this case we omit term $\xi\mathcal{K}D(\gamma-i\Omega)$ from the numerator and $|D|^2\mathcal{K}$ from the denominator of last term in \eqref{Sdetb}. The expression for the noise power spectral density transforms to
\begin{equation}
    \begin{split}
      S(\Omega)&=2\gamma_m(n_T+1) +\frac{\gamma_m^2+\Omega^2}{\mathcal{K}}+\\
    &+\mathcal{K} \left(\frac{\delta^2}{\gamma^2+\Omega^2}+\frac{4\gamma^2\Delta^2}{(\gamma^2+\Omega^2)^2}\right) 
 \label{ssmall}   
    \end{split}
\end{equation}
Here the last term describes the residual back action because of the frequency detunings. 
 
We find the optimal pump parameter $\mathcal{K}_{opt}$ that minimizes \eqref{ssmall} and present it in form
\begin{align}
   \mathcal{K}_{opt}=\frac{\sqrt{\gamma_m^2+\Omega^2}(\gamma^2+\Omega^2)}{\sqrt{\delta^2(\gamma^2+\Omega^2)+4\gamma^2\Delta^2}}.
\end{align}
Comparison of $\mathcal{K}_{opt}$ with $\mathcal{K}_{crit1}$ shows that in order to satisfy \eqref{cond_small} the detunings have to obey to the condition
\begin{equation}
     \gamma^2\Delta^2 \ll \delta^2(\delta^2 +4 \Delta^2).
\end{equation}
This is impossible if $|\Delta| \geq |\delta|$ since $|\delta|<\gamma$, per our initial assumption. Hence, in this case we can drop the back action term and the noise power spectral density \eqref{ssmall} becomes
 \begin{equation}
     S(\Omega) =2\gamma_m(n_T+1) +\frac{\gamma_m^2+\Omega^2}{\mathcal{K}},
 \end{equation}
 It reaches its minimum 
\begin{align} 
     S(\Omega)&=2\gamma_m(n_T+1)+S_{SQL,f}\frac{|\Delta|}{2|\delta|}.\label{sminsmall} 
 \end{align} 
when  $\mathcal{K} \approx \mathcal{K}_{crit1}$.

In case $|\Delta| \ll |\delta|$ we find
\begin{align} 
     S(\Omega)&=2\gamma_m(n_T+1)+S_{SQL,f}\frac{|\delta|}{\sqrt{\gamma^2+\Omega^2}}. \label{sminsmall2}
 \end{align} 

For the case of higher power pump
\begin{align}
 \mathcal {K}_{crit2}=\frac{\sqrt{\gamma^2+\Omega^2}\sqrt{\gamma_m^2+\Omega^2}}{\Delta}\gg \mathcal {K}\gg\mathcal{K}_{crit1} \label{cond_avg}
 \end{align}
we omit term $\delta$ from the numerator and $|D|^2\mathcal{K}$ from the denominator of last term in \eqref{Sdetb}. The noise power spectral density takes form
 \begin{align} 
 S(\Omega)&=2\gamma_m(n_T+1) +\frac{\gamma_m^2+\Omega^2}{\mathcal{K}}+ \mathcal K^3|D|^2. \label{savg}
\end{align}
The optimal pump parameter $\mathcal{K}_{opt}$, that minimizes \eqref{savg}, is
\begin{equation}
    K_{opt}=\frac{(\gamma_m^2+\Omega^2)^{1/2}(\gamma^2+\Omega^2)^{1/4}}{3^{1/4}|\Delta|^{1/2}}.
\end{equation}
Comparison of $\mathcal{K}_{opt}$ with  $\mathcal{K}_{crit1}$ and $\mathcal{K}_{crit2}$ shows that in order to satisfy \eqref{cond_avg} the detunings have to follow the relation
\begin{align}
    1 \gg    |\Delta|/\gamma   \gg \delta^2/\gamma^2, 
\end{align} 
which is feasible. The minimal noise power spectral density in this case equals to
\begin{subequations}
\begin{align}
\begin{split}
&S_{min} =2\gamma_m(n_T+1)+\\
 &+\left(3^{1/4}+\frac{1}{3^{1/4}}\right)\frac{(\gamma_m^2+\Omega^2)^{1/2}\Delta^{1/2}}{(\gamma^2+\Omega^2)^{1/4}}=
\end{split}\\ 
& =2\gamma_m(n_T+1)+\frac{\sqrt{3}+1}{2\sqrt[4]{3}} \frac{\Delta^{1/2}}{(\gamma^2+\Omega^2)^{1/4}}S_{SQL,f}. 
\end{align}\label{sminavg}
\end{subequations}
 
Finally, for the large pump power
%
\begin{align}
  \mathcal K  |D| &\gg 1, 
 \quad \text{or}\ \mathcal {K}\gg \mathcal {K}_{crit2}.\label{cond_big}
 \end{align}
%
we omit term $\delta$ from the numerator  and $1$ from the denominator of last term in \eqref{Sdetb}. Noise power spectral density takes form
 \begin{equation}
     \begin{split}
       S(\Omega)&=2\gamma_m(n_T+1) + \mathcal K
    +\frac{4\gamma^2(\gamma_m^2+\Omega^2)}{\mathcal K(\gamma^2+\Omega^2)}. \label{sbig}  
     \end{split}
 \end{equation} 
The optimal pump parameter $\mathcal{K}_{opt}$ minimizing \eqref{sbig} becomes
\begin{equation}
 K_{opt}=\sqrt{\frac{4\gamma^2(\gamma_m^2+\Omega^2)}{\mathcal (\gamma^2+\Omega^2)}}.   
\end{equation} 
Comparison of $\mathcal{K}_{opt}$ with   $\mathcal{K}_{crit2}$ shows that in order to satisfy \eqref{cond_big} the detunings have to follow the relationship $1 \ll  (2  \Delta )/\gamma  $, contradicting our assumption that $\Delta\ll \gamma$. Therefore, in this case the back action term proportional to the pump power dominates and the noise power spectral density \eqref{sbig}  becomes
 \begin{equation}
     S(\Omega) \simeq 2\gamma_m(n_T+1) +\mathcal{K}.
 \end{equation}
It reaches minimum at $\mathcal{K}=\mathcal{K}_{crit2}$ and it's minimum is related to SQL (\ref{SQL}, \ref{SSQL}) as
\begin{subequations}
\begin{align} \nonumber
     S(\Omega)&=2\gamma_m(n_T+1) + \mathcal{K}_{crit2}=\\
     &=2\gamma_m(n_T+1)+S_{SQL,f}\frac{\sqrt{\gamma^2+\Omega^2}}{2\Delta}. 
 \end{align} \label{sminbig}
\end{subequations}
Comparing the results (\ref{sminsmall}, \ref{sminsmall2}, \ref{sminavg}, \ref{sminbig}) we find that the regime of the intermediate pump power provides the minimal noise spectral density for the case of $|\Delta| \gg |\delta|$, while for the opposite case, $|\Delta| \ll |\delta|$, the limit of smaller power is optimal.

\section{Coherent coupling and quantum mechanics-free subsystems}

In this section we discuss in detail the fundamental features of the scheme proposed here that lead to the back-action evasion.

\subsection{Coherent coupling}
It is possible to argue that our measurement technique realizes coherent coupling, proposed in \cite{LiPRA2019}, between the optical modes and the mechanical mode. Unlike the traditional dispersive coupling, in our case the mechanical displacement does not affect the frequencies of the optical modes. Instead, it rotates the basis vectors of amplitude distribution coefficients.

Ring resonator with a partially reflective mirror \cite{LiPRA2019} is an example of coherent coupling. The mirror opens the degeneracy of the initial clockwise and anticlockwise modes in this resonator and creates the new symmetric and antisymmetric eigenmodes with the point of node and antinode on the input mirror. Displacement $x$ of this mirror shifts the position of this point by $x$, which can be considered as the rotation of the basis vectors representing the eigenmodes, without change of the eigenfrequencies. 

Let us start from the Hamiltonian of the scheme, written in the matrix form. 
\begin{align}
& H=\hbar \left(\hat c_0\ \hat c_+\ \hat c_-\right)^\dag \begin{pmatrix}
\omega_0& i\eta^* \hat d^\dag & -i\eta \hat d\\
 -i\eta \hat d& \omega_+ &0 \\
i \eta^* \hat d^\dag &0&  \omega_-
\end{pmatrix}\begin{pmatrix}
\hat c_0\\
\hat c_+\\
\hat c_-
\end{pmatrix} 
\end{align} 
Considering the mechanical mode operator $\hat d$ as a parameter, we diagonalize the matrix and find the eigenfrequencies of the system.
\begin{subequations}
\begin{align}
&(\omega_0- \lambda)(\omega_+-\lambda)(\omega_--\lambda)-2|\eta d|^2(\omega -\lambda) =0\\
&(\omega_0-\lambda)\left((\omega_\lambda)^2-\omega_m^2-2|\eta d|^2\right)=0,\\
&\lambda_1=\omega_0,\\
&\lambda_{2,3}=\omega\pm\sqrt{\omega_m^2+2|\eta \hat d|^2}\approx \omega\pm \omega_m = \omega_\pm
\end{align}  
\end{subequations} 
In the linear approximation the eigenfrequencies $\lambda_{1,2,3}$ do not depend on the mechanical degree of freedom $d$. Eigenmodes $\hat c_{1,2,3}$, corresponding to the eigenfrequencies $\lambda_{1,2,3}$, in linear approximation, can be expressed via initial (partial) modes as 
\begin{subequations}
\begin{align*}
 \hat c_1 &=\left(1,\ \frac{i\eta  \hat d }{\omega_m},\ \frac{ i\eta^* \hat d^\dag }{\omega_m}\right)\begin{pmatrix}
\hat c_0\\
\hat c_+\\
\hat c_-
\end{pmatrix} ,& {\bf v}_1&=\left(1,\ \frac{i\eta  \hat d }{\omega_m},\ \frac{ i\eta^* \hat d^\dag }{\omega_m}\right),\\
 \hat c_2 &=\left(\frac{ i\eta^*  \hat d^\dag }{\omega_m},\ 1,\ 0\right)\begin{pmatrix}
\hat c_0\\
\hat c_+\\
\hat c_-
\end{pmatrix} ,& {\bf v}_2&=\left(\frac{ i\eta^*  \hat d^\dag }{\omega_m},\ 1,\ 0\right),\\
\hat c_3 &=\left(\frac{ i\eta   \hat d   }{\omega_m},\ 0,\ 1\right)\begin{pmatrix}
\hat c_0\\
\hat c_+\\
\hat c_-
\end{pmatrix} ,& {\bf v}_3&=\left(\frac{ i\eta   \hat d   }{\omega_m},\ 0,\ 1\right).
\end{align*}  
  \end{subequations} 

This expression can be explained in terms of the coherent coupling concept. If $\hat d=0$ (the mechanical oscillator is in the equilibrium), the eigenmodes transform into the initial optical modes $\hat c_1 \rightarrow\hat c$, $\hat c_2 \rightarrow \hat c_+$, $\hat c_3 \rightarrow \hat c_-$, which allows us to use them in our analysis.

${\bf v}_{1,2,3}$ are the vectors of amplitude distribution coefficients. In linear approximation they are orthogonal and have constant norm equal to 1: $({\bf v}_i,{\bf v}_j)=\delta_{ij}+O(\hat d^2)$. Thus the coupling between optical and mechanical modes does not change the lengths of the basis vectors rotating them instead.

Let us compare our scheme with the similar scheme proposed earlier \cite{21a1VyNaMaPRA}. It is based on   the Michelson-Sagnac interferometer with a partially transparent mirror. It also has two nondegenerate optical modes and the movement of the mirror provides the coupling between them.   In that scheme we analyze it's Hamiltonian (again we consider the mechanical mode operator $\hat d$ as a parameter).
\begin{align}
& H=\hbar \left( \hat c_+\ \hat c_-\right)^\dag \begin{pmatrix}
\omega_+ &-i\eta \hat d \\
i\eta^* \hat d^\dag&  \omega_-
\end{pmatrix}\begin{pmatrix}
\hat c_+\\
\hat c_-
\end{pmatrix} 
\end{align} 
The eigenfrequencies of this system are
 \begin{align}
\lambda_{1,2}=\omega_++\omega_-\pm\sqrt{(\omega_+-\omega_-)^2+4|\eta \hat d|^2}\approx \omega_\pm
\end{align}    
The corresponding eigenmodes can be expressed via initial modes as 
\begin{subequations}
\begin{align}
\hat c_1 &=\left(1,\ \frac{i\eta \hat d }{\omega_m}\right)\begin{pmatrix} 
\hat c_+\\
\hat c_-
\end{pmatrix} ,\\
\hat c_2 &=\left(\frac{ i\eta^*  \hat d^\dag }{\omega_m},\ 1\right)\begin{pmatrix} 
\hat c_+\\
\hat c_-
\end{pmatrix} .
\end{align}  
  \end{subequations}  

Therefore,  this system also represents the coherent coupling. The difference between the schemes is in the interaction structure. It the scheme proposed here the optical sidebands $c_\pm$ do not interact with each other directly, instead the interaction goes on via the central mode $c$. Moreover, the eigenmode of the sideband $c_2$ (or $c_3$)  does not depend on partial mode of the respective opposite sideband $c_-$ (or $c_+$).

In the scheme from \cite{21a1VyNaMaPRA} there is no intermediate mode, so the two modes have to interact with each other. It leads to an instability due to the ponderomotive nonlinearity. Our scheme is free of the instability, thanks to the presence and of the central mode $c$. 

\subsection{QMFS and back-action evasion}

To explain how back-action evasion is realized in our scheme we  present our system in terms of the quantum mechanics-free subsystems (QMFS), introduced in \cite{TsangPRX2012}. A set of variables $\{X_1,...X_n\}$ forms a  QMFS  if
\begin{equation}
    \forall i,j,\, \forall t,t'\, [X_i(t),X_j(t')]=0.
\end{equation}
In this case the measurement of a variable $X_i$ at  time $t$ does not perturb any of the variables $X_j$ ($i\ne j$) from the set and they can be precisely measured at time $t'$.

Since the quantities that we observe in the experiment are quadrature amplitudes, to identify independent QMFSs for our system we have to present the Hamiltonian in terms of the observables. We provide a simplified description of the procedure in this section, while the strict and detailed derivation can be found in Appendix \ref{QMFSHam}.

We start from the equations of motion \eqref{quadIn} for the quadrature amplitudes and remove decay and pump, which corresponds to the analysis of a closed system.  
\begin{subequations}
 \begin{align}  
\dot c_{+a} &=  -\eta C_0  d_a ,\\ 
\dot c_{+\phi} &= -  \eta C_0  d_\phi  ,\\
\dot  c_{-a} &= \eta C_0  d_a  ,\\
\dot c_{-\phi} &=- \eta  C_0  d_\phi ,\\
\dot  d_a &= \eta C_0 \Big(c_{+a}+c_{-a}\Big),\\
\dot  d_\phi & = \eta C_0  \Big(c_{+\phi} - c_{-\phi}\Big).
 \end{align}
 \end{subequations}

These equations of motion are generated by the Hamiltonian
\begin{equation}
    V=\hbar \eta C_0 (c_{+a}+c_{-a}) d_a + \hbar \eta C_0 ( c_{+\phi} -c_{-\phi} ) d_\phi.
\end{equation}
(it coincides with \eqref{VappB} derived in Appendix \ref{QMFSHam}).
We introduce
\begin{align}
 d_a&= Q, & d_\phi&=P,\\
 \frac{c_{+a}+c_{-a}}{\sqrt2}&=\Phi_1, & \frac{c_{+\phi}+c_{-\phi}}{\sqrt2}&=\Pi_1,\\
 \frac{c_{+a}-c_{-a}}{\sqrt2}&=\Phi_2, & \frac{c_{+\phi}-c_{-\phi}}{\sqrt2}&=\Pi_2.
\end{align} 
Operators $Q$ and $P$, as well as $\Phi_{1,2}$ and $\Pi_{1,2}$, are quantum conjugated, that is
\begin{equation}
[Q,P]=[\Phi_1,\Pi_1]=[\Phi_2,\Pi_2]= i \delta_{jk}.
\end{equation} 
The other variables of the system commute with each other. We can rewrite the Hamiltonian as
\begin{align} 
V&= \sqrt{2}\hbar | \eta C| \Phi_1 Q  +\sqrt2 \hbar | \eta C|\Pi_2 P.
\end{align}  
The equation of motions are
\begin{align}
 \dot \Pi_1&= \sqrt{2}  | \eta C|  Q, & \dot Q&=\sqrt2   | \eta C|\Pi_2, & \dot \Pi_2&=0\\
 \dot \Phi_2& = \sqrt2  | \eta C|  P, &  \dot P&=-\sqrt{2}  | \eta C| \Phi_1,& \dot \Phi_1& = 0.
\end{align}

Their solution in the time domain is 
\begin{align}
\left\{
\begin{aligned}
\Pi_1&=\Pi_{10}+ \sqrt{2}  | \eta C| Q_0t+    | \eta C|^2\Pi_{20}t^2,\\
Q_{\phantom{0}}&= Q_0+\sqrt2   | \eta C|\Pi_{20}t,\\
\Pi_2&=\Pi_{20},
\end{aligned}\right. \label{qmfs1} \\
\left\{
\begin{aligned}
\Phi_2 &=\Phi_{20}+ \sqrt2  | \eta C|   P_0t-  | \eta C|^2 \Phi_{10}t^2,\\
P_{\phantom{0}}&=P_0-\sqrt{2}  | \eta C| \Phi_{10}t,\\
\Phi_1 &= \Phi_{10}.
\end{aligned}\right.  \label{qmfs2}
\end{align}
Here $X_0=X(0)$ for each variable $X(t)$.

As we can see, every variable from this system is QND-variable. It is due to the fact that none of them depends dynamically from their conjugate. Moreover, all of the variables from the upper \eqref{qmfs1} (or lower \eqref{qmfs2}) set  commute with each other. That is why they form the QMFS. Therefore, we have two independent (in the dynamic sense) QMFSs: $\{\Pi_1,\,Q,\,\Pi_2\}$ and $\{\Phi_2,\,P,\,\Phi_1\}$.

Let us consider the subsystem  $\{\Pi_1,\,Q,\,\Pi_2\}$ \eqref{qmfs1}. The meaning  of each term in the equation for $\Pi_1$ is as follows. $\Pi_{10}$ corresponds to shot noise, $\sqrt{2}  | \eta C| Q_0t$ corresponds to the signal and $ | \eta C|^2\Pi_{20}t^2$ is the back-action. We recall that $ \Pi_1=\cfrac{c_{+\phi}+c_{-\phi}}{\sqrt2}$, and  $ \Pi_2=\cfrac{c_{+\phi}-c_{-\phi}}{\sqrt2}$. It happens because this system has two outputs and two phase quadrature amplitudes of the output fields, corresponding to   $c_p$ and $c_m$. We can independently take their sum ($\Pi_1$) and difference ($\Pi_2$) and perform a quantum demolition measurement of $Q=d_a$, which would allow us to get the information about the signal force acting on that quadrature.

We have removed the decay and pump terms and considered the closed system in the analysis presented above, to explain how the QMFSs appear in this scheme. The analysis of the realistic scheme presented in Sec. \ref{Solution} is in full agreement with this consideration.

It is interestig to compare our scheme with the measurement scheme based on ``negative mass'' by Tsang and Caves (TC) \cite{TsangPRX2012}. The Hamiltonian of their general model is 
\begin{equation}
    V_{TC}=\frac{m \omega^2 q^2}{2}+\frac{p^2}{2m}-\frac{m \omega^2 q'^2}{2}-\frac{p'^2}{2m}.
\end{equation}

The observable variables that correspond to real physical systems are  $\{q,\,p\}$ (for example, coordinate and momentum of a mechanical oscillator) and $\{q',\,p'\}$ (for example, quadratures of the auxiliary optical resonator). The force acts on $q$. None of these variables is QND. The QND variables are
\begin{align}
 q+q'&= Q_{TC}, & \frac{p+p'}{2}&=P_{TC},\\
 \frac{q-q'}{2}&=\Phi_{TC}, &  p-p' &=\Pi_{TC}.
\end{align} 

Measuring $Q_{TC}$ would allow one to get the information about the signal force and avoid the back-action.

The scheme of the broadband variation measurement proposed here has several distinguished features:

{\em a)} all of its variables are already of QND nature;\

{\em b)} these QND variables $Q$, $P$ and $\Phi_{1,2}$, $\Pi_{1,2}$ correspond to parameters of real physical systems ($Q$ and $P$ are the quadrature amplitudes of the mechanical oscillator, $\Phi_{1,2}$ and $\Pi_{1,2}$ are quadrature amplitudes of the optical modes);\

{\em c)} there are three degrees of freedom in our scheme (two optical and one mechanical), whereas in scheme \cite{TsangPRX2012} only two degrees of freedom are considered. The presence of the three degrees of freedom enables measurements of the optical quadratures in the two channels and results in the cancellation of the back action in a broad band;\

{\em d)} the variables $\Phi_{1,2}$, $\Pi_{1,2}$ correspond to the probes, while in scheme \cite{TsangPRX2012} measurement of $Q_{TC}$ has to be made with an additional probe.

\section{Discussion}

In this paper we have introduce a broadband back action evading measurement of a classical mechanical force. In the measurement scheme described by Fig.~\ref{scheme} the signal information contained in the mechanical {\em quadratures} transfers to the optical quadratures. The measurement of the difference of the optical amplitude quadratures is equivalent to the registration of the mechanical  amplitude quadrature, whereas the measurement of the sum of the optical phase quadratures  corresponds to the registration of mechanical  phase quadrature, as shown by Eq.~\eqref{quadIn}. This is a peculiar property of the parametric interaction. 

One of the main features of the proposed here measurement strategy is in the usage of the single probe field with detection in the two independent quantum outputs. It gives us a flexibility to measure back action separately and then subtract it completely from the measurement result. The subtraction of back action can be made in a broad frequency band. 

In contrast, in conventional variational measurements \cite{93a1VyMaJETP, 95a1VyZuPLA, 02a1KiLeMaThVyPRD} there is only one quantum output and the back action cannot be measured separately from the signal. Measurement of the linear combination of the amplitude and phase quadratures in that case allows partial subtraction of the back action. Only one quadrature of the output wave has to be measured to surpass the SQL.

The scheme proposed here allows a measurement either of a combination of sum and difference amplitude quadratures \eqref{beta-a2}  or sum and difference of phase  quadratures \eqref{betaMSIphi}. Generalization \eqref{quadphi} is also possible. These measurement lead to back action evasion in a broad frequency band. We expect that this technique will find a realization in other configurations.

Our study represents the further development of the broadband dichromatic variational measurement \cite{21a1VyNaMaPRA}. The main advantages of the current study include:
(i) Our scheme uses one optical pump, in contrast with the two pumps considered in \cite{21a1VyNaMaPRA}. (ii) The configuration proposed in \cite{21a1VyNaMaPRA} requires compensation of the resonant classical mechanical force impinged by the dichromatic pump on the mechanical oscillator.  Our scheme is free of it. (iii) Proposed here configuration is free from parasitic back action which takes place in scheme  \cite{21a1VyNaMaPRA}

It worth noting that a similar idea was recently considered in electro-optical configuration \cite{22JOSABNaMaVy}. However, in that case the classical force depended on the attenuation of the radio frequency system, while in our case there is no such dependence. Additionally, here we have considered a nonideal case involving various frequency detunings and found the validity range of the technique.
\begin{figure}
 \includegraphics[width=0.3\textwidth]{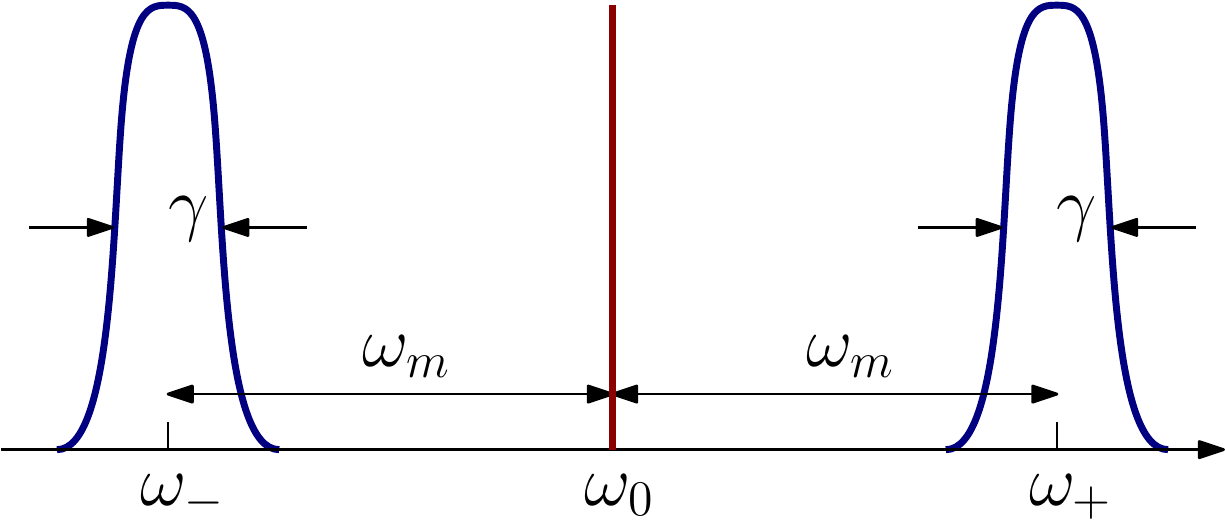}
 \caption{Two modes scheme.}\label{scheme22}
\end{figure}

In this paper we have considered three mode scheme, as shown in Fig.~\ref{scheme}. We can use {\em two} modes scheme with frequencies $\omega_0\pm \omega_m$ and use the pump with frequency $\omega_0$ located in between of the modes --- see Fig.~\ref{scheme22}. The back action can be suppressed in this scheme as well. However, more optical power will be needed to beat the SQL.

\section{Conclusion}

We have shown that the simultaneous and independent measurements of optimal quadrature amplitudes of two optical harmonics generated due to ponderomotive interaction of light and a mechanical force mediated by an opto-mechanical interaction enables a back action evading measurement. The back action can be removed from the signal by postprocessing of the measurement data. The measurement becomes feasible since the opto-mechanical system is an example of the quantum configuration containing quantum-mechanics-free subsystems lending themselves for continuous quantum nondemolition measurements \cite{TsangPRX2012}. 

We hope that proposed here broadband coherent multidimensional variational measurement can be used in precision optomechanical measurements including laser gravitational wave detectors.

 \acknowledgments
The research of SPV and AIN has been supported by the Russian Foundation for Basic Research (Grant No. 19-29-11003), the Interdisciplinary Scientific and Educational School of Moscow University ``Fundamental and Applied Space Research'' and from the TAPIR GIFT MSU Support of the California Institute of Technology. AIN is the recipient of a Theoretical Physics and Mathematics Advancement Foundation “BASIS” scholarship (Contract No. 21-2-10-47-1). The reported here research performed by ABM was carried out at the Jet Propulsion Laboratory, California Institute of Technology, under a contract with the National Aeronautics and Space Administration (80NM0018D0004). This document has LIGO number P2200131.  

\appendix

\section{Derivation of the intracavity fields}
\label{IntrDeriv}
 
In this appendix we provide details of the standard calculation for intracavity fields, for example, see \cite{Walls2008}.

We begin with Hamiltonian \eqref{Halt}
\begin{align}  
   \label{Ht}
      H_T &= \sum\limits_{k=0}^\infty \hslash \omega_k b_k^\dag b_k,\\
      H_\gamma & =i \hslash \sqrt\frac{\gamma \Delta \omega}{\pi} \sum\limits_{k=0}^\infty \left[(c_+^\dag+c_-^\dag) b_k-(c_++c_-)b_k^\dag \right],\nonumber\\
       \label{Htm}
      &H_{T, \, m} = \sum\limits_{k=0}^\infty \hslash \omega_k q_k^\dag q_k,\\
      \label{Hgammam}
      &H_{\gamma_m}  =i \hslash \sqrt\frac{\gamma \Delta \omega}{\pi} \sum\limits_{k=0}^\infty \left[d^\dag q_k-dq_k^\dag \right].
  \end{align}
Here $H_T$ is the Hamiltonian of the environment presented as a bath of  oscillators described with frequencies $\omega_k=\omega_{k-1}+\Delta \omega$ and annihilation and creation operators $b_k$, $b_k^\dag$.  $H_\gamma$ is the Hamiltonian of coupling between the environment and the optical resonator, $\gamma$ is the coupling constant. Similarly $H_{T, \, m}$ is the Hamiltonian of the environment presented by a thermal bath of mechanical oscillators with frequencies $\omega_k=\omega_{k-1}+\Delta \omega$ and amplitudes described with annihilation and creation operators $q_k$, $q_k^\dag$.  $H_{\gamma_m}$ is the Hamiltonian of coupling between the environment and the mechanical oscillator, $2\gamma_m$ is the decay rate of the oscillator.
  
We write Heisenberg equations for operators $c_+$ and $b_k$:
\begin{subequations} 
  \begin{align}
i\hslash \dot c_+ & =[c_+,H]=\hslash\omega_+ c_+-i\hslash \eta c_- d+i \hslash \sqrt\frac{\gamma \Delta \omega}{\pi} \sum\limits_{k=0}^\infty b_k,\\
i \hslash \dot b_k & =[b_k,H]=\hslash \omega_k b_k-i \hslash \sqrt\frac{\gamma \Delta \omega}{\pi}  \left(c_++c_-\right)
  \end{align} \label{Heis1}
  \end{subequations}

We introduce slow amplitudes $c_\pm \rightarrow c_\pm e^{-i \omega_\pm t}$, $d \rightarrow d e^{-i(\omega_+-\omega_-)t}$, $b_k \rightarrow b_k e^{-i \omega_k t}$ and substitute them into \eqref{Heis1}
\begin{subequations} 
  \begin{align}
\dot c_+ & =-\eta c_- d+  \sqrt\frac{\gamma \Delta \omega}{\pi} \sum\limits_{k=0}^\infty b_k e^{-i(\omega_k-\omega_+)t}, \label{Heisc}\\
  \dot b_k & = -\sqrt\frac{\gamma \Delta \omega}{\pi} \left(c_+e^{-i(\omega_+-\omega_k)t}+c_- e^{-i(\omega_--\omega_k)t}\right) \label{Heisb}
  \end{align}
  \end{subequations}

Using initial condition $b_k(t=0)=b_k(0)$ to integrate \eqref{Heisb} we derive
\begin{align}
b_k(t) & =b_k(0)-\int\limits_0^t\sqrt\frac{\gamma \Delta \omega}{\pi}  c_+(s)e^{-i(\omega_+-\omega_k)s}ds-\\
& -\int\limits_0^t\sqrt\frac{\gamma \Delta \omega}{\pi}c_-(s)e^{-i(\omega_--\omega_k)s}ds \label{binit}
\end{align} 
Using the condition $b_k(t=\infty)=b_k(\infty)$ to integrate \eqref{Heisb} we derive
\begin{align}
b_k(t) & =b_k(\infty)+\int\limits_t^\infty\sqrt\frac{\gamma \Delta \omega}{\pi}  c_+(s)e^{-i(\omega_+-\omega_k)s}ds-\\
& +\int\limits_t^\infty\sqrt\frac{\gamma \Delta \omega}{\pi}c_-(s)e^{-i(\omega_--\omega_k)s}ds \label{bfin}
\end{align} 
To get the input-output relation we substitute initial condition \eqref{binit} into \eqref{Heisc}
\begin{subequations} 
  \begin{align}
\dot c_+ & =-\eta c_- d+   \sum\limits_{k=0}^\infty \sqrt\frac{\gamma \Delta \omega}{\pi} b_k(0)e^{-i(\omega_k-\omega_+)t}-\\
&-\sum\limits_{k=0}^\infty\int\limits_0^t\frac{\gamma \Delta \omega}{\pi}  c_+(s) 
	 e^{-i(\omega_k-\omega_+)(t-s)}ds -\\
&-\left(\sum\limits_{k=0}^\infty\int\limits_0^t\frac{\gamma \Delta \omega}{\pi}  c_-(s) e^{-i(\omega_k-\omega_-)(t-s)}ds \right)e^{i(\omega_+-\omega_-)t}\nonumber
  \end{align} \label{Heis2}
  \end{subequations}
and omit the last term proportional to $e^{i(\omega_+-\omega_-)t}$ as the fast oscillating, and define the input field
  \begin{equation}
a_+(t) = \sum\limits_{k=0}^\infty \sqrt\frac{ \Delta \omega}{2\pi} b_k(0)e^{-i(\omega_k-\omega_+)t} \label{aplus}
\end{equation}
To calculate the remaining sum in \eqref{Heis2} we  assume the validity of limit $\Delta \omega \rightarrow 0$ and  replace the sum by the integration
\begin{subequations} 
  \begin{align} 
\Delta \omega &\sum\limits_{k=0}^\infty \rightarrow \int\limits_0^\infty d \omega_k\\
&\sum\limits_{k=0}^\infty\int\limits_0^t\frac{\gamma \Delta \omega}{\pi}  c_+(s) e^{-i(\omega_k-\omega_+)(t-s)}ds  \rightarrow\\ \nonumber
\rightarrow & \int\limits_0^\infty\int\limits_0^t 2\gamma  c_+(s) e^{-i(\omega_k-\omega_+)(t-s)}ds \frac{d \omega_k}{2 \pi}=\\ \nonumber
= &  \int\limits_{-\omega_+}^\infty\int\limits_0^t 2\gamma  c_+(s) e^{-i \omega(t-s)}ds \frac{d \omega}{2 \pi}\approx\\ \nonumber
\approx &  \int\limits_{-\infty}^\infty\int\limits_0^t 2\gamma  c_+(s) e^{-i \omega(t-s)}ds \frac{d \omega}{2 \pi}=\\
= & \int\limits_0^t 2\gamma  c_+(s) \delta(t-s) ds = \frac{2 \gamma c_+(t)}{2}=\gamma c_+(t)
  \end{align} \label{gammac}
  \end{subequations}

Substituting \eqref{aplus} and \eqref{gammac} into \eqref{Heis2} we obtain
   \begin{align}
\dot c_+ & =-\eta c_- d+  \sqrt{2\gamma}a_+-\gamma c_+\\
\dot c_+ & + \gamma c_++\eta c_- d=  \sqrt{2\gamma}a_+. \label{cplus}
\end{align}

By analogue, we derive the equation for input field $a_-$ and present it in a similar form
  \begin{equation}
a_-(t) = \sum\limits_{k=0}^\infty \sqrt\frac{ \Delta \omega}{2\pi} b_k(0)e^{-i(\omega_k-\omega_-)t}
\end{equation}
It leads to the equation for the intracavity field $c_-$
   \begin{align}
\dot c_-+ \gamma c_--\eta c_+ d^\dag=  \sqrt{2\gamma}a_-. \label{cminus}
\end{align}

Similar equation can be derived for the amplitude $q(t)$ of the mechanical oscillator
  \begin{equation}
q(t)=  \sum\limits_{k=0}^\infty \sqrt\frac{ \Delta \omega}{2\pi} b_{m, \,k}(0)e^{-i(\omega_k-\omega_+)t},
\end{equation}
resulting in the Langevin  equation for mechanical oscillator quadrature $d$
   \begin{align}
\dot d+ \gamma_m d-\eta^* c_+ c_-^\dag=  \sqrt{2\gamma_m}q.
\end{align}
To derive the output relation we substitute \eqref{bfin} into \eqref{Heisc} and define the output fields 
  \begin{align}
b_+(t) = -\sum\limits_{k=0}^\infty \sqrt\frac{ \Delta \omega}{2\pi} b_k(\infty)e^{-i(\omega_k-\omega_+)t} \\
b_-(t) = -\sum\limits_{k=0}^\infty \sqrt\frac{ \Delta \omega}{2\pi} b_k(\infty)e^{-i(\omega_k-\omega_+)t}.
\end{align}
It leads to 
   \begin{align} 
\dot c_+ & - \gamma c_++\eta c_- d=  \sqrt{2\gamma}b_+ \label{bplus}\\
\dot c_- & - \gamma c_-+\eta^* c_+ d^\dag=  \sqrt{2\gamma}b_- \label{bminus}
\end{align}
Utilizing pairs of equations \eqref{cplus} and \eqref{bplus}  as well as \eqref{cminus} and \eqref{bminus} we obtain the final expression for the input-output relations
\begin{align}
b_+=-a_++\sqrt{2\gamma} c_+\\
b_-=-a_-+\sqrt{2\gamma} c_-
\end{align}

Let us to derive the commutation relations for the Fourier amplitudes of the operators. We introduce Fourier transform of field $a_+(t)$ using \eqref{aplus}
 \begin{subequations}
 \begin{align}
a_+(\Omega) = \int\limits_{-\infty}^{\infty} \sum\limits_{k=0}^\infty \sqrt\frac{ \Delta \omega}{2\pi} b_k(0)e^{-i(\omega_k-\omega_+-\Omega)t}dt=\\
= \sum\limits_{k=0}^\infty \sqrt { 2 \pi \Delta \omega}  b_k(0)  \delta(\Omega-\omega_k+\omega_+)
 \end{align}
 \end{subequations}
 This allows us to find the commutators \eqref{comm1}: 
  \begin{subequations}
 \begin{align} \nonumber
 [a_+(\Omega)& ,a_+^\dag(\Omega') ] 
   =\sum\limits_{k=0}^\infty  2 \pi \Delta \omega   [b_k(0),b_k^\dag(0)] \times\\ \nonumber
&\times \delta(\Omega-\omega_k+\omega_+)  \delta(\Omega'-\omega_k+\omega_+) \rightarrow\\ \nonumber
&\rightarrow \int\limits_{-\infty}^\infty 2 \pi [b(0),b^\dag(0)]  \delta(\Omega-\omega)  \delta(\Omega'-\omega) d\omega=\\
&=2\pi \delta(\Omega-\Omega'),
 \end{align}
 \end{subequations}
 and the correlators \eqref{corr1}
   \begin{subequations}
 \begin{align} \nonumber
\langle a_+(\Omega)&,a_+^\dag(\Omega') \rangle
 = \sum\limits_{k=0}^\infty  2 \pi \Delta \omega   \langle b_k(0),b_k^\dag(0) \rangle \times\\ \nonumber
&\times \delta(\Omega-\omega_k+\omega_+)  \delta(\Omega'-\omega_k+\omega_+) \rightarrow\\ \nonumber
&\rightarrow \int\limits_{-\infty}^\infty 2 \pi \langle b(0),b^\dag(0) \rangle \delta(\Omega-\omega)  \delta(\Omega'-\omega) d\omega=\\
&=2\pi \delta(\Omega-\Omega')
 \end{align}
 \end{subequations}
Similar expressions can be derived for commutators and correlators of the optical $a_-$ and mechanical $q$ quantum amplitudes.

\section{Hamiltonian presented using quadrature amplitudes} \label{QMFSHam}
 
We start from the full Hamiltonian describing interaction of modes of a lossless nonlinear cavity. We assume that all of the fields (denoted by hats) depend on time
\begin{subequations}
\begin{align} 
H_0&=\hbar \omega \hat c^\dag \hat c+\hbar \omega_- \hat c_-^\dag \hat c_- +\hbar \omega_+ \hat c_+^\dag \hat c_++\hbar \omega_m\hat  d^\dag \hat d,\\
\label{Va}
V&=\hbar \eta(\hat c_-^\dag \hat d^\dag +\hat c_+^\dag \hat d)\hat c+\hbar \eta^* (\hat c_- \hat d  +\hat c_+ \hat d^\dag)\hat c^\dag.
\end{align}  
  \end{subequations} 
 
From the analysis made earlier it is known, that small fluctuations of $\hat c$ mode do not influence the system. We change it to mean field $\hat c \rightarrow C$ and omit the term in $H_0$ connected to it. For all of the other modes the resonator is closed. 
  
We express the creation and annihilation operators via their corresponding quadratures:
  \begin{subequations}
\begin{align}
&\hat c_\pm=\frac{\hat c_{\pm a}+i \hat c_{\pm \phi}}{\sqrt2}, \qquad  \hat d=\frac{\hat d_a +i \hat d_\phi }{\sqrt2},\\
&\hat c_\pm^\dag=\frac{\hat c_{\pm a}-i \hat c_{\pm \phi}}{\sqrt2}, \qquad  \hat d^\dag=\frac{\hat d_a -i \hat d_\phi }{\sqrt2}.
\end{align}
\end{subequations}
  
The Hamiltonian transforms to
\begin{subequations}
\begin{align} 
\begin{split}
    &H_0=\frac{\hbar \omega_-}{2}(\hat c_{-a}^2 +\hat c_{-\phi}^2)+\frac{\hbar \omega_+}{2}(\hat c_{+a}^2 +  \hat c_{+\phi}^2)+\\
&+\frac{\hbar \omega_m}{2}(\hat d_a^2+\hat d_\phi^2),
\end{split}\\
\label{Vb}
&V= \frac{\hbar (\eta C+\eta^* C^*)}{2} \left((\hat c_{-a}+\hat c_{+a}) \hat d_a+(\hat c_{+\phi}-\hat c_{-\phi})\hat d_\phi\right)+\\
&+\frac{i\hbar (\eta^* C^*-\eta C)}{2}\left(  (\hat c_{-\phi}+\hat c_{+\phi}) \hat d_a+ (\hat c_{-a}-\hat c_{+a})  \hat d_\phi \right).\nonumber
\end{align}  
  \end{subequations} 

We assume that the coupling constant  is real $\eta=\eta^*$ and that the mean amplitude of the intracavity field is real $C_0=C_0^*$. Then $\eta C=  \eta C_0 e^{-i \omega t }$, and the   Hamiltonian   transforms to
\begin{subequations}
\begin{align} 
\begin{split}
    &H_0=\frac{\hbar \omega_-}{2}(\hat c_{-a}^2 +\hat c_{-\phi}^2)+\frac{\hbar \omega_+}{2}(\hat c_{+a}^2 +  \hat c_{+\phi}^2)+\\
&+\frac{\hbar \omega_m}{2}(\hat d_a^2+\hat d_\phi^2),
\end{split}\\
V&= \hbar  \eta C_0  ((\hat c_{+ a}+\hat c_{- a})\cos{\omega t}-(\hat c_{+\phi}+\hat c_{-\phi})\sin{\omega t})  \hat d_a+\nonumber\\
&+\hbar  \eta C_0((\hat c_{+\phi}-\hat c_{-\phi})\cos{\omega t}+(\hat c_{+ a}-\hat c_{- a})\sin{\omega t}) \hat d_\phi.\nonumber
\end{align}  
  \end{subequations} 

The Hamiltonian $H_0$ corresponds to the free evolution of the quadratures. It is instructive to introduce slow amplitudes 
\begin{subequations}
\begin{align}
&\hat c_{\pm a}= c_{\pm a} \cos{\omega_\pm t}+c_{\pm\phi}\sin{\omega_\pm t},\\
&\hat c_{\pm\phi}= c_{\pm\phi} \cos{\omega_\pm t}-c_{\pm a}\sin{\omega_\pm t},\\
&\hat d_{ a}= d_{ a} \cos{\omega_m t}+d_{ \phi}\sin{\omega_m t},\\
&\hat d_{ \phi}= d_{ \phi} \cos{\omega_m t}-d_{  a}\sin{\omega_m t}.
\end{align}
\end{subequations}
Using simple arithmetic we arrive at
\begin{subequations}
\begin{align} \nonumber
&\hat c_{\pm a}\cos{\omega t}- \hat c_{\pm\phi} \sin{\omega t}=c_{\pm a}\cos{\omega_m t}\pm c_{\pm\phi} \sin{ \omega_m t},\\
&\hat c_{\pm\phi}\cos{\omega t}+ \hat c_{\pm a} \sin{\omega t}=c_{\pm\phi}\cos{\omega_m t} \mp c_{\pm a} \sin{ \omega_m t}. \nonumber
\end{align}
\end{subequations}

Substituting these expressions into the interaction Hamiltonian $V$ we get
\begin{align}
&V=\hbar  \eta C_0 (c_{+a}+c_{-a}) d_a+\hbar  \eta C_0(c_{+\phi}-c_{-\phi})  d_{\phi}. \label{VappB}
\end{align}
%


\end{document}